\documentclass[a4paper,12pt]{article}
\usepackage{graphicx} 

\usepackage{amsfonts, amsmath, amssymb, array, color, enumerate, epsf, graphicx, grffile, tikz,  cancel, subfig, float, ifthen, xcolor, amsthm, dsfont}

\renewcommand{\title}[1]{\vbox{\center\LARGE{#1}}\vspace{5mm}}
\renewcommand{\author}[1]{\vbox{\center#1}\vspace{5mm}}

\newcommand{\email}[1]{\vbox{\center\tt#1}\vspace{5mm}}

\usepackage[utf8]{inputenc}
\usepackage[T1]{fontenc}

\usepackage{color}
\usepackage{epsfig}

\usepackage{bm,amsfonts,mathrsfs}
\numberwithin{equation}{section}

\usepackage[a4paper, total={6in, 9in}]{geometry}

\usepackage{url}
\usepackage[numbers]{natbib}
\usepackage[
      colorlinks=true,
      linkcolor=black,
      urlcolor=blue,
      filecolor=black,
      citecolor=red,
      ]{hyperref}

\newcommand{\tr}{\mathrm{Tr}}

\begin{document}

\begin{titlepage}

\begin{center}

\hfill{\;}

\vspace{2cm}

{\Large\bf Konishi lifts a black hole}

\vspace{1.5cm}

\author{Jaehyeok Choi$^{1}$ and Eunwoo Lee$^{2}$}

\vspace{0.7cm}

\textit{$^1$School of Physics, Korea Institute for Advanced Study,\\
85 Hoegi-ro, Dongdaemun-gu, Seoul 02455, Republic of Korea}\\

\vspace{0.2cm}

\textit{${}^2$Department of Theoretical Physics, \\ Tata Institute
	of Fundamental Research, Homi Bhabha Rd, Mumbai 400005, India}

\vspace{0.7cm}

\email{jaehyeokchoi@kias.re.kr,  eunwoo.lee@tifr.res.in}

\end{center}

\vspace{1cm}

\begin{abstract}
We investigate the \emph{quantum} cohomology of a supercharge $Q$ in $\mathcal{N}=4$ super Yang-Mills theory.
Recent analyses have revealed a mismatch between the one-loop BPS spectra of the S-dual $SO(7)$ and $Sp(3)$ theories.
The $SO(7)$ theory contains a pair of additional graviton (monotone) and non-graviton (fortuitous) cohomologies, whose net contributions cancel in the superconformal index.
We show that the quantum-corrected $Q$, inferred from the generalized Konishi anomaly, pairs and lifts these extra cohomologies.
\end{abstract}

\vfill

\end{titlepage}

\tableofcontents

\section{Introduction}
Investigating the microstates of supersymmetric black holes in anti-de Sitter (AdS) space from the dual superconformal field theories (SCFTs) \cite{Maldacena:1997re} has been highly fruitful \cite{Grant:2008sk,Chang:2013fba,Chang:2022mjp,Choi:2022caq,Choi:2023znd,Budzik:2023vtr,Chang:2023zqk,Choi:2023vdm,Chang:2024zqi,Chang:2024lxt,deMelloKoch:2024pcs,Chang:2025rqy,Gadde:2025yoa,Chang:2025mqp,Chen:2025sum,Kim:2025vup}, especially following the successes in reproducing black hole degeneracies \cite{Cabo-Bizet:2018ehj,Choi:2018hmj,Benini:2018ywd} from the superconformal index \cite{Kinney:2005ej,Romelsberger:2005eg}.
Specifically, the problem is formulated in terms of the cohomology of a supercharge $Q$.
Finding Bogomol’nyi-Prasad-Sommerfield (BPS) states of the SCFT on $\mathbb{R}\times S^3$ is equivalent to finding BPS operators in $\mathbb{R}^4$ via the state-operator correspondence. And by the standard Hodge-theoretic argument, such BPS operators (annihilated by $Q$ and $Q^\dag$) are in one-to-one correspondence with $Q$-cohomology classes.

The cohomologies (cohomology classes) corresponding to the (BPS) Kaluza-Klein (multi-)particles in the dual AdS gravitational theory are called \emph{gravitons}. Or, since they are $Q$-closed for any $N$, they are also called \emph{monotone} \cite{Chang:2024zqi}.
Non-graviton cohomologies (linearly independent of gravitons) are sometimes referred to as \emph{black hole} cohomologies. Or, based on the fact that they become $Q$-closed only after using finite $N$ trace relations, they are also called \emph{fortuitous} \cite{Chang:2024zqi}.

Due to the strong-weak nature of the AdS/CFT duality, computations of $Q$-cohomology have necessarily relied on the conjecture that the one-loop BPS spectrum of $\mathcal{N}=4$ super Yang-Mills (SYM) theory is exact \cite{Grant:2008sk,Chang:2022mjp}.\footnote{
For less supersymmetric theories without known Lagrangians, the situation is less well understood \cite{LSscft,conifold}.
}
One-loop BPS operators correspond to tree level (`half-loop') cohomologies.\footnote{
Here, the term `$n$-loop' denotes the order $\sim g_{YM}^{2n}$ corrections to the classical Dilatation operator \cite{Beisert:2004ry}.
Following this convention, the classical $Q$-cohomology is often referred to as `one-loop' cohomology. However, we will continue to refer to classical $Q$-cohomology as tree level or classical.
}
Based on this non-renormalization conjecture, most works have studied only classical $Q$-cohomology, even though the aim is to investigate strong-coupling microstates, except for the works on holomorphic twist \cite{Budzik:2023xbr,Budzik:2022mpd,Bomans:2023mkd,Gaiotto:2024gii}.

Recently, evidence against the one-loop exactness of the BPS spectrum was found by \cite{Chang:2025mqp}, based on \cite{Gadde:2025yoa} that identified a fortuitous cohomology in the so-called BMN sector of $SO(7)$ $\mathcal{N}=4$ SYM. 
It was found that the classical $Q$-cohomology is different for $SO(7)$ and $Sp(3)$ theories, while they are believed to be the same theory with different coupling constants related by $\tau\to -1/\tau$
where $\tau=\frac{\theta}{2\pi}+\frac{4\pi i}{g_{YM}^2}$.
Specifically, it was reported \cite{Chang:2025mqp} that there is a pair of additional graviton $O_m$ and black hole $O_f$ cohomologies in the $SO(7)$ case.
\begin{table}[H]
    \centering
    \begin{tabular}{|c||c|c|}
         \hline
         cohomologies & fortuitous $O_f$ & monotone $O_m$ \\ \hline \hline
         $(J_1,J_2,R_1,R_2,R_3)$ & $(\frac{1}{2},\frac{1}{2},\frac{5}{2},\frac{5}{2},\frac{5}{2})$ & $(0,0,3,3,3)$ \\ \hline
         number of letters & 8 & 8 \\ \hline
    \end{tabular}
    \caption{The pair of additional \emph{classical} $Q$-cohomology classes in $SO(7)$ theory that are absent in $Sp(3)$ classical $Q$-cohomology. $(J_1,J_2)$ denote the Cartans of $SO(4)$ that rotates orthogonal 2-planes of $\mathbb{R}^4$, and $(R_1,R_2,R_3)$ denote the Cartans of $SO(6)_R$ rotating orthogonal 2-planes of $\mathbb{R}^6$.}
    \label{tab:The pair of additional cohomologies}
\end{table}
\vspace{-0.5cm}
If $S$-duality holds, it is natural to expect that the two additional classical cohomologies are paired and lifted once quantum effects are taken into account. In particular, the difference in the charges of these two cohomologies is precisely that carried by the supercharge $Q=Q^4_-$ of $\mathcal{N}=4$ SYM.
And indeed, the result of this paper is that the quantum-corrected $Q$-action on $O_f$ yields $O_m$.

More precisely, we find that $QO_f=O_m$ holds \emph{within} the classical cohomology.
Let $Q_0$ be the classical supercharge and $Q_n$ its $n$-loop correction.
Then the cohomology of the full quantum supercharge $Q=Q_0+Q_1+\cdots$ can be obtained iteratively, by computing the cohomology of $Q_n$ \emph{restricted to} $Q_{n-1}$-cohomology $H^{\bullet}(H^{\bullet}(\cdots,Q_{n-1}),Q_{n})$ \cite{Budzik:2023xbr}. The full quantum $Q$ could contain the non-perturbative contributions as well. The fact that this iteratively obtained cohomology is equivalent to the $Q$-cohomology is explained in Appendix D of \cite{Budzik:2023xbr}, using the notion of homotopy transfer.
In our case, it suffices to consider the one-loop correction $Q_1$ only.
Then, the $Q$-cohomology is equivalent to the $\tilde{Q}_1$-cohomology, where $\tilde{Q}_1=\pi_0 Q_1 \iota_0$ is the restriction of $Q_1$ to the classical cohomology. Here, $\pi_0$ is the projection to $Q_0$-cohomology, and $\iota_0$ is the inclusion such that $\pi_0\iota_0$ is the identity map on the $Q_0$-cohomology. Roughly speaking, this equivalence (`quasi-isomorphism') is because $Q_0$ acts trivially on the $Q_0$-cohomology.


A notable feature of the extra graviton operator $O_m$ is that it belongs to the (anti) chiral ring sector, preserving $1/8$ of supersymmetries. 
It is a Lorentz singlet and vanishes on the classical Coulomb branch, where one can choose a gauge such that the scalars and gauginos lie in the Cartan subalgebra of the gauge algebra.

More generally, let us call operators that remain nonzero when all adjoint fields are restricted to the Cartan subalgebra \emph{Coulomb-type}, and those that vanish under this restriction \emph{non-Coulomb-type}.
Coulomb-type operators are never classically $Q$-exact, since the tree level $Q$-action always produces a commutator. Hence, Coulomb-type gravitons are necessarily non-trivial in cohomology.
Moreover, there exists a natural one-to-one correspondence between the Coulomb-type gravitons of $SO(2N+1)$ and those of $Sp(N)$ \cite{Gadde:2025yoa}, as their Cartan subalgebras are related by a similarity transformation (up to a direct sum with a $1\times 1$ zero matrix to match dimensions) \cite{Chang:2025mqp}.
Consequently, the extra graviton $O_m$ of $SO(7)$ must be of non-Coulomb type, explaining its vanishing on the classical Coulomb branch.
Our result is that the classically non-exact $O_m$ is quantum mechanically $Q$-exact, implying that its vacuum expectation value also vanishes at the quantum level.
This, in particular, confirms the validity of counting chiral ring operators as $N$ identical particles in a three-dimensional bosonic and two-dimensional fermionic harmonic oscillator \cite{Kinney:2005ej}, which had appeared to be challenged by the existence of the classically non-Coulomb chiral ring operator $O_m$.

Since all Coulomb gravitons naturally survive under S-duality, it is plausible that any graviton lifted by quantum corrections must necessarily be of the non-Coulomb type.
Moreover, the non-Coulomb nature of a graviton depends on the rank of the gauge group, in a manner analogous to the $Q$-closedness of fortuitous cohomologies: as the rank increases, a non-Coulomb graviton eventually becomes Coulomb-type.
In our case, the non-Coulomb graviton $O_m$ becomes Coulomb-type when the gauge group is $SO(9)$, consistent with the fact that the extra fortuitous operator $O_f$ of $SO(7)$ becomes non-closed for $SO(9)$ \cite{Gadde:2025yoa}.

It is worth noting that the argument for the one-loop exactness is based on the assumption that $Q$ acts as a differential operator satisfying the Leibniz rule and that it defines an associative cohomology \cite{Chang:2022mjp}. Therefore, it is expected that quantum effects violate these assumptions.

A natural source of such a quantum effect is the Konishi anomaly. As we briefly review in Section~\ref{section 2}, the Konishi anomaly is the supersymmetrized version of the Adler–Bell–Jackiw (ABJ) anomaly. It acts as a second-order differential operator and therefore fails to satisfy the Leibniz rule, breaking one of the key assumptions of the argument.
It has often been believed that the Konishi anomaly does not affect the BPS spectrum in $\mathcal{N}=4$ SYM—in the sense that it does not lift any classical BPS cohomologies. In this note, however, we show that this is not the case: the Konishi anomaly plays a crucial role by lifting the fortuitous cohomology and pairing it with a monotone state to form a long multiplet.

This note is structured as follows. In Section~\ref{section 2}, we review the Konishi anomaly and its generalization. In Section~\ref{section 3}, we show that the Konishi anomaly lifts the $SO(7)$ fortuitous cohomology by pairing it with a monotone state. We conclude in Section~\ref{section 4} with discussions and remarks.

\section{Generalized Konishi anomaly and quantum $Q$}\label{section 2}

In this section, we briefly review the generalized Konishi anomaly, and infer the quantum correction of $Q$ from it.

The Konishi anomaly is the supersymmetric analogue of the ABJ anomaly, arising as the anomalous divergence of the classically conserved (or non-conserved when the superpotential is non-vanishing) Konishi current superfield associated with the phase rotation of chiral multiplet superfields \cite{Konishi:1983hf,Konishi:1985tu}.
This is distinct from the $U(1)_R$, because it does not act on the superspace coordinate, and is broken classically by the presence of a superpotential.

In $\mathcal{N}=1$ gauge theories with chiral multiplet $\Phi$ in the adjoint representation, it takes the form
\begin{align}\label{kon}
    \bar{D}^2 \tr \bar{\Phi}e^{[V,\cdot]}\Phi\Big|_{anomaly}=g_{YM}^2\frac{C(\mathrm{adj})}{16\pi^2}\tr W_\alpha W^{\alpha}\;.
\end{align}
where $W_\alpha=\lambda_\alpha+\cdots$ is the gaugino superfield.
The right-hand side can be regarded as the supersymmetric counterpart of the gauge field strength contribution in the ABJ anomaly.
When the superpotential $W$ is non-vanishing, there also exists a classical contribution that is proportional to $\tr \Phi \frac{\partial W}{\partial \Phi}$.
The trace is taken over the fundamental representation (when the gauge group is $SU(N)$ or $Sp(N)$), or the vector representation ($SO(N)$).
Also, we defined $C(R)$ as the dual Coxeter number of the representation $R$.
$$\tr T_R^A T_R^B=C(R)\delta^{AB}\;.$$

A generalization of the Konishi anomaly, applicable to any composite operator of the form $\tr \bar{\Phi}e^{[V,\cdot]}f(\Phi,W_\alpha)$, was presented in \cite{Cachazo:2002ry} (see also the review \cite{Argurio:2003ym}).
For a general adjoint chiral multiplet $\Phi^A$ with a simple gauge group $G$,
\begin{align}\label{generalized Konishi}
    \bar{D}^2 \bar{\Phi}^A (e^{[V,\cdot]}\Phi)^B\Big|_{anomaly}
    =\frac{g_{YM}^2}{16\pi^2}(W_{\alpha}^{\mathrm{adj}}W^{\mathrm{adj}\,\alpha})^{AB}\;,
\end{align}
where $A,B$ are adjoint indices, and $W^{\mathrm{adj}}_\alpha=W_\alpha^A (T^A_{\mathrm{adj}})$.
From this, it is straightforward to reproduce \eqref{kon}.
Note that the equation \eqref{generalized Konishi} is applicable when all the other operator insertions are chiral, i.e. functions of $\Phi,W_\alpha$.

When the gauge group is $SU(N)$ or $U(N)$, for example, the equation \eqref{generalized Konishi} can be rewritten as
\begin{equation}\label{kon su}
\begin{aligned}
    \frac{16\pi^2}{g_{YM}^2} \bar{D}^2 \bar{\Phi}^i\!_j (e^{[V,\cdot]}\Phi)^k\!_l\Big|_{anomaly}
    &=-C(\mathrm{fund})\tr [e_j\!^i,W_\alpha][e_l\!^k,W^\alpha]\\
    &=\frac{1}{2}\big((W_\alpha W^\alpha)^i\!_l\delta^k_j 
    +(W_\alpha W^\alpha)^k\!_j\delta^i_l
    -2(W_\alpha)^i\!_l (W^\alpha)^k\!_j \big)\;,
\end{aligned}
\end{equation}
where we used $C(\mathrm{fund})=1/2$ and $(e_j\!^i)^k\!_l:=\delta_{j}^k\delta_{l}^i$.

When the gauge group is $SO(N)$, which is the relevant case for our discussion, one finds
\begin{align}\label{kon so}
    \frac{16\pi^2}{g_{YM}^2} \bar{D}^2 \bar{\Phi}_{ij} (e^{[V,\cdot]}\Phi)_{kl}\Big|_{anomaly}
    &=-C(\mathrm{vec})\frac{1}{4}\tr [e_{ji}-e_{ij},W_\alpha][e_{lk}-e_{kl},W^\alpha]\nonumber\\
    &=\frac{1}{2}\big(
    \delta_{il}(W_\alpha W^\alpha)_{jk}+\delta_{jk}(W_\alpha W^{\alpha})_{il}
    -\delta_{jl}(W_\alpha W^\alpha)_{ik}-\delta_{ik}(W_\alpha W^\alpha)_{jl}
    \big)\nonumber\\
    &\quad +(W_\alpha)_{il} (W^\alpha)_{jk}-(W_\alpha)_{jl}(W^\alpha)_{ik}
\end{align}
where $C(\mathrm{vec})=1$ and $(e_{ij})_{kl}:=\delta_{ik}\delta_{jl}$.
Taking $k=j,l=i$, one reproduces 
$$\bar{D}^2\tr \bar{\Phi}e^{[V,\cdot]}\Phi\Big|_{anomaly}
=g_{YM}^2\frac{(N-2)}{16\pi^2}\tr W_\alpha W^\alpha\;,$$ correctly ($C(\mathrm{adj})=N-2$ for $SO(N)$). 

From this, one can infer a one-loop correction to the $Q$-action, which we denote as $Q_{K}$:
\begin{align}\label{eq:QK}
    Q_{K}=\bar{A}_{ij,kl}
    \frac{g_{YM}^2}{32\pi^2}
    (\partial_\psi)_{ij}(\partial_{\bar\Phi})_{kl},
\end{align}
where
\begin{equation}
\begin{aligned}
    A_{ij,kl}&=\delta_{il}(\lambda_\alpha \lambda^\alpha)_{jk}
    +\delta_{jk}(\lambda_\alpha \lambda^\alpha)_{il}
    +2 (\lambda_\alpha)_{il}(\lambda^\alpha)_{jk}\\
    &-\delta_{jl}(\lambda_\alpha \lambda^\alpha)_{ik}
    -\delta_{ik}(\lambda_\alpha \lambda^\alpha)_{jl}
    -2(\lambda_\alpha)_{jl}(\lambda^\alpha)_{ik}\;.
\end{aligned}
\end{equation}
and $(\partial_\psi)_{ij}:=\frac{\partial}{\partial(\psi_+)_{ij}}$, $(\partial_{\bar\Phi})_{kl}:=\frac{\partial}{\partial \bar{\Phi}_{kl}}$.
Note that there is a bar on $A_{ij,kl}$ in the equation \eqref{eq:QK}, which means that one should replace $\lambda_\alpha$ with $\bar{\lambda}_{\dot\alpha}$.
Also, we defined $\psi_+=Q_+\Phi$, which is a BPS letter for the supercharge $Q=Q_-$.
From this, for example,
\begin{align*}
    Q\tr \bar{\Phi}\psi_+ = g_{YM}\tr\bar\Phi \frac{\partial \bar{W}}{\partial \bar\Phi}+  g_{YM}^2\frac{(N-2)}{16\pi^2}\tr \bar{\lambda}_{\dot\alpha} \bar{\lambda}^{\dot\alpha}\;,
\end{align*}
The first term is the contribution from the classical $Q_0$-action, while the second term is from the one-loop correction $Q_K$.
Since $Q_{K}$ is a second derivative, it does not preserve the number of traces.
It can increase or decrease the number of traces by one.

For more complicated composite operators, there could be other quantum corrections to the supercharge, even at the same $g_{YM}^2$ order.
For example, when there are BPS derivatives on $\psi_+$ or $\Phi$, the differential operator \eqref{eq:QK} must also contain terms that differentiate $D_{+\dot\alpha}\psi_+$ or $D_{+\dot\alpha}\Phi$.

Finally, since we are going to apply this to the $Q=Q^4_-$ cohomology of $\mathcal{N}=4$ SYM, the derivatives in \eqref{eq:QK} should be understood as 
$$\frac{\partial^2}{\partial (\psi_{+m})_{ij}\partial(\bar{\phi}^m)_{kl}}$$
where $\bar{\phi}^m,\psi_{+m}$ ($m=1,2,3$) are the BPS letters of $Q=Q^4_-$.
For more details of our convention, see, for example, table 1 of \cite{Choi:2023znd}.

\section{Konishi lifts a black hole}\label{section 3}

In this section, we show that the Konishi anomaly lifts a classical fortuitous cohomology by forming a long multiplet with a monotone state.

It was reported in \cite{Chang:2025mqp} that in $SO(7)$ $\mathcal{N}=4$ SYM, there exists a monotone non-Coulomb cohomology $O_{m}$ in the (anti) chiral ring sector with charges 
$$(J_1,J_2,R_1,R_2,R_3)=(0,0,3,3,3)$$
and $8$ letters. This bosonic operator's contribution to the superconformal index is canceled by a fortuitous cohomology $O_f$ found in \cite{Gadde:2025yoa}, with charges  
$$(J_1,J_2,R_1,R_2,R_3)=\big(\frac{1}{2},\frac{1}{2},\frac{5}{2},\frac{5}{2},\frac{5}{2}\big)$$
Both of these (tree-level) cohomologies are not found in the classical cohomology of the S-dual theory; that is, the $Sp(3)$ $\mathcal{N}=4$ SYM \cite{Chang:2025mqp}.

Let us find an explicit representative of $O_m$.
Since it should be an $SU(3)_R$ singlet, there are not many possibilities.
The letter content is restricted to $(\bar{\lambda}_{\dot\alpha})^2(\bar\phi^m)^6$.
For the ease of notation, let us denote the anti-chiral fields $\bar{\lambda}_{\dot\alpha}$ and $\bar{\phi}^m$ as $\lambda_\alpha$ and $\phi^m$. 
Also, since it was reported to be a graviton (=monotone) \cite{Chang:2025mqp}, $O_m$ should be a multi-trace of
\begin{equation}
\begin{aligned}
    u^{mn}:=\tr \phi^m\phi^n\;,\;\;\;
    S:=\tr \lambda_\alpha\lambda^\alpha\;,\;\;\;
    S^m_\alpha:=\tr \phi^m\lambda_\alpha\;,\;\;\;
    S^{mn}:=\tr \lambda_\alpha\lambda^\alpha \phi^{(m}\phi^{n)}\;,\;
\end{aligned}
\end{equation}
We have not included the monotone operators with an odd number of letters in the trace, as they vanish identically. This is because the matrices are antisymmetric and the chiral ring condition requires the scalars and gauginos to commute with each other (anti-commute for gaugino-gaugino). 
Also, we do not have to consider the chiral ring operators with more than 2 scalars. The chiral ring condition forces the fields inside the trace to be (anti) symmetric; therefore, the operators are always in a totally symmetric representation of $SU(3)_R$. When there are 3 or more scalars inside a trace, there is no way to form an $SU(3)_R$ singlet in the charge sector we want. This is because one has to use precisely two epsilon tensors $\epsilon_{mnp}$ of $SU(3)_R$ to form a singlet with the letters $(\lambda_\alpha)^2(\phi^m)^6$.
Then, we see that there are only 3 operators.
\begin{equation}
\begin{aligned}
    M_1&:=\epsilon_{mnp}\epsilon_{qrs}(u^{mq}S)u^{nr}u^{ps}\;,\\
    M_2&:=\epsilon_{mnp}\epsilon_{qrs}
    (S^{ m}_{\alpha}S^{q\alpha })u^{nr}u^{ps}\;,\\
    M_3&:=\epsilon_{mnp}\epsilon_{qrs}S^{mq}u^{nr}u^{ps}\;,
\end{aligned}
\end{equation}
These three operators are the same when the fields are in the Cartan subalgebra. Since we are considering Lorentz scalar operators in the chiral ring, it can be rephrased that they are the same on the Coulomb branch.
\begin{align*}
    M_1=3M_2=6M_3\quad \text{(on Coulomb branch)}
\end{align*}
Namely, there are two non-Coulomb operators, and according to \cite{Chang:2025mqp}, one of the two non-Coulomb operators is $Q_0$-exact, while the other is not.

We obtain the $Q_0$-exact combination, given by
\begin{align}
M_1 + 2M_2 - 10M_3.
\end{align}
Therefore, any linear combinations of $M_1$, $M_2$, and $M_3$ that are linearly independent of this $Q_0$-exact operator correspond to monotone states.
And one of them is of Coulomb-type, while the other is of non-Coulomb-type.

Now we are going to act $Q_{K}$ to the ($Q_0$-closed) fortuitous operator $O_f$, which is given as \cite{Gadde:2025yoa}
\begin{align}
    &O_f=\text{Tr}[Y^2] \text{Tr}[X\psi_1] \text{Tr}[XZ]^2-4 \text{Tr}[Y^2] \text{Tr}[XZ]
   \text{Tr}[ZXZ\psi_3]-\text{Tr}[XZ]^2
   \text{Tr}[ZY\psi_3Y]\nonumber\\
   &-4 \text{Tr}[XZ]^2
   \text{Tr}[ZY^2\psi_3]+8 \text{Tr}[XZ] \text{Tr}[ZXY^2Z\psi_3]+4
   \text{Tr}[XZ] \text{Tr}[ZXZY\psi_3Y]\nonumber\\
   &+16 \text{Tr}[XZ]
   \text{Tr}[ZXZY^2\psi_3]-4 \text{Tr}[Z\psi_3] \text{Tr}[ZX]
   \text{Tr}[ZXY^2]+8 \text{Tr}[ZXZ\psi_3] \text{Tr}[XYZY]\nonumber\\
   &-2 \text{Tr}[Y^2]
   \text{Tr}[Z\psi_3] \text{Tr}[ZXZX]+8 \text{Tr}[ZXZX]
   \text{Tr}[Y^2Z\psi_3]+2 \text{Tr}[ZXZX] \text{Tr}[YZY\psi_3]\nonumber\\
   &+16\text{Tr}[YZ\psi_3] \text{Tr}[ZXZXY]+8 \text{Tr}[Z\psi_3]
   \text{Tr}[ZXZXY^2]+8 \text{Tr}[Y^2] \text{Tr}[ZXZXZ\psi_3]\nonumber\\
   &+16\text{Tr}[ZXZXYZY\psi_3]-8 \text{Tr}[ZXZXZY\psi_3Y]-32
   \text{Tr}[ZXZXZY^2\psi_3]-16 \text{Tr}[ZXZYXYZ\psi_3]\nonumber\\
   &-16\text{Tr}[ZXZYXZY\psi_3]-16 \text{Tr}[ZXZY^2XZ\psi_3]-(grav)
\end{align}
Here, $-(grav)$ indicates that we project the state onto the subspace orthogonal to the monotone state vector space. For further details, see \cite{Gadde:2025yoa}.

Since both $O_f$ and $Q_{K}$ are singlets under the $SU(3)$ flavor symmetry, the anomalous contribution $Q_KO_f$ is also an $SU(3)$ singlet (up to $Q_0$-exact terms). Also, $Q_kO_f$ consists of $(\lambda_{\alpha})^2(\phi^m)^6$. Consequently, it can be expressed in terms of $M_1$, $M_2$, and $M_3$, up to $Q$-exact terms.

Using equation \eqref{kon so}, one finds that $QO_f=Q_{K}O_f$ is proportional to $M_1 - M_2 - 4M_3\propto O_m$, up to $Q_0$-exact terms. Namely,
$$Q_K O_f = O_m +Q_0\Lambda$$
for some $\Lambda$.
As briefly explained in the introduction, $O_m$ being $Q_K$-exact in the $Q_0$-cohomology (i.e. up to $Q_0$-exactness) is equivalent to $O_m$ being $Q$-exact in the full space of operators.
We checked this by explicitly constructing the $Q_0$-exact basis in the charge sector $J_a=0,R_i=3$ with 8 letters, which consists of 741 operators \cite{Chang:2025mqp}.
The combination $M_1-M_2-4M_3$ is linearly independent of the 741 $Q_0$-exact operators, and $Q_KO_f$ is a linear combination of $M_1-M_2-4M_3$ and $Q_0$-exact operators.
Note that this combination $M_1-M_2-4M_3$ is non-Coulomb. Namely, it vanishes when the fields are in the Cartan subalgebra.
Since this combination belongs to the classical $Q_0$-cohomology, it was a monotone state before the Konishi anomaly was taken into account. However, once the quantum correction is included, this monotone state pairs with the fortuitous cohomology to form a long multiplet.

\section{Discussion}\label{section 4}

In this paper, we have shown that the quantum-corrected supercharge can lift (tree level) fortuitous and monotone states, as demonstrated in the example of the $\mathcal{N}=4$ $SO(7)$ gauge theory.
While we inferred the one-loop correction to the supercharge from the generalized Konishi anomaly, which is sufficient for our purposes, there can be other quantum corrections for more complicated composite operators.
The quantum correction of the supercharge essentially arises from the renormalization of composite operators due to interactions.
In our case, it was due to the presence of the Yukawa coupling term $\sim \int d^4x \tr \phi_m\{\bar{\lambda}^{\dot\alpha},\bar{\psi}_{\dot\alpha}^m\}$.
It would be interesting to study what kind of quantum corrections could arise from the other interaction terms.


As noted in the introduction, the non-Coulomb monotone cohomologies exhibit behavior analogous to that of the fortuitous cohomologies: as $N$ increases, they become Coulomb-type, just as the fortuitous cohomologies lose their $Q$-closedness.
Also, (as explained in the introduction) since the $SO(2N+1)$ Coulomb cohomologies are naturally mapped to the same cohomologies in the S-dual $Sp(N)$, if a monotone is lifted, it is plausible that it is non-Coulomb. From this, it seems natural that the fortuitous and non-Coulomb monotone cohomologies are lifted together.
Of course, other combinations, such as (non-Coulomb, non-Coulomb) or (fortuitous, fortuitous), could also be lifted. 
And the lifting of a Coulomb monotone cohomology is logically possible.
If one could argue that $Q_K$ maps non-Coulomb to non-Coulomb, since all known fortuitous cohomologies in $\mathcal{N}=4$ are non-Coulomb, the lifting of Coulomb monotone should occur by pairing it with another Coulomb monotone.
It would be interesting to find examples of such cases.


$\mathcal{N}=4$ SYM classical cohomology admits a consistent truncation to the so-called BMN sector, which discards the BPS letters with dotted Lorentz spinor indices $(\bar{\lambda}_{\dot\alpha},D_{+\dot\alpha})$.
In the $SO(7)/Sp(3)$ duality, the BMN indices were found to be different \cite{Gadde:2025yoa}, which predicts infinitely many (classical) fortuitous or non-Coulomb graviton cohomologies (which are conformal primaries) in either $SO(7)$ or $Sp(3)$ theory. This is because the spectra of the Coulomb BMN gravitons are isomorphic for $SO(7)$ and $Sp(3)$.
It would be interesting to explain this discrepancy by quantum $Q$, although it is not clear whether the predicted fortuitous/non-Coulomb monotone should be lifted quantum mechanically or not.


One can ask whether the first fortuitous operator found in $SU(2)$ SYM \cite{Chang:2022mjp,Choi:2022caq} is lifted due to the Konishi anomaly.
Since it has a representative that contains the letters $\bar{\phi}^m,\psi_{n+}$ only \cite{Choi:2023znd}, we can use $Q_K$ for the one-loop correction.
For the $SU(2)$ threshold operator, we find that it is not lifted by the Konishi anomaly: the action of $Q_{K}$ on the operator vanishes up to $Q_0$-exact terms.\footnote{See related discussions in \cite{Chang:2025mqp} indicating that the threshold state remains unlifted under the general quantum effect.}

It would also be interesting to explore non-perturbative corrections of $Q$ in $\mathcal{N}=4$ SYM. 
Also, the quantum corrected $Q$-cohomology in more general gauge theories with fewer supersymmetries \cite{seiberg} is another direction to pursue. 

\section*{Acknowledgments}
We would like to thank Abhijit Gadde, Seok Kim and Shiraz Minwalla for useful comments and discussions. 
The work of J.C. was supported by a KIAS Individual Grant PG106601
at the Korea Institute for Advanced Study. The work of E.L. was supported by the Infosys Endowment for the study of the Quantum Structure of Spacetime.

\bibliography{References}

@article{Maldacena:1997re,
    author = "Maldacena, Juan Martin",
    title = "{The Large $N$ limit of superconformal field theories and supergravity}",
    eprint = "hep-th/9711200",
    archivePrefix = "arXiv",
    reportNumber = "HUTP-97-A097, HUTP-98-A097",
    doi = "10.4310/ATMP.1998.v2.n2.a1",
    journal = "Adv. Theor. Math. Phys.",
    volume = "2",
    pages = "231--252",
    year = "1998"
}

@article{Grant:2008sk,
    author = "Grant, Lars and Grassi, Pietro A. and Kim, Seok and Minwalla, Shiraz",
    title = "{Comments on 1/16 BPS Quantum States and Classical Configurations}",
    eprint = "0803.4183",
    archivePrefix = "arXiv",
    primaryClass = "hep-th",
    reportNumber = "IMPERIAL-TP-08-SK-01",
    doi = "10.1088/1126-6708/2008/05/049",
    journal = "JHEP",
    volume = "05",
    pages = "049",
    year = "2008"
}

@article{Chang:2013fba,
    author = "Chang, Chi-Ming and Yin, Xi",
    title = "{1/16 BPS states in $\mathcal N=$ 4 super-Yang-Mills theory}",
    eprint = "1305.6314",
    archivePrefix = "arXiv",
    primaryClass = "hep-th",
    doi = "10.1103/PhysRevD.88.106005",
    journal = "Phys. Rev. D",
    volume = "88",
    number = "10",
    pages = "106005",
    year = "2013"
}

@article{Chang:2022mjp,
    author = "Chang, Chi-Ming and Lin, Ying-Hsuan",
    title = "{Words to describe a black hole}",
    eprint = "2209.06728",
    archivePrefix = "arXiv",
    primaryClass = "hep-th",
    doi = "10.1007/JHEP02(2023)109",
    journal = "JHEP",
    volume = "02",
    pages = "109",
    year = "2023"
}

@article{Choi:2022caq,
    author = "Choi, Sunjin and Kim, Seok and Lee, Eunwoo and Park, Jaemo",
    title = "{The shape of non-graviton operators for SU(2)}",
    eprint = "2209.12696",
    archivePrefix = "arXiv",
    primaryClass = "hep-th",
    reportNumber = "KIAS-P22052",
    doi = "10.1007/JHEP09(2024)029",
    journal = "JHEP",
    volume = "09",
    pages = "029",
    year = "2024"
}

@article{Choi:2023znd,
    author = "Choi, Sunjin and Kim, Seok and Lee, Eunwoo and Lee, Siyul and Park, Jaemo",
    title = "{Towards quantum black hole microstates}",
    eprint = "2304.10155",
    archivePrefix = "arXiv",
    primaryClass = "hep-th",
    doi = "10.1007/JHEP11(2023)175",
    journal = "JHEP",
    volume = "11",
    pages = "175",
    year = "2023",
    note = "[Erratum: JHEP 03, 091 (2025)]"
}

@article{Budzik:2023vtr,
    author = "Budzik, Kasia and Murali, Harish and Vieira, Pedro",
    title = "{Following Black Hole States}",
    eprint = "2306.04693",
    archivePrefix = "arXiv",
    primaryClass = "hep-th",
    month = "6",
    year = "2023"
}

@article{Chang:2023zqk,
    author = "Chang, Chi-Ming and Feng, Li and Lin, Ying-Hsuan and Tao, Yi-Xiao",
    title = "{Decoding stringy near-supersymmetric black holes}",
    eprint = "2306.04673",
    archivePrefix = "arXiv",
    primaryClass = "hep-th",
    doi = "10.21468/SciPostPhys.16.4.109",
    journal = "SciPost Phys.",
    volume = "16",
    number = "4",
    pages = "109",
    year = "2024"
}

@article{Choi:2023vdm,
    author = "Choi, Jaehyeok and Choi, Sunjin and Kim, Seok and Lee, Jehyun and Lee, Siyul",
    title = "{Finite N black hole cohomologies}",
    eprint = "2312.16443",
    archivePrefix = "arXiv",
    primaryClass = "hep-th",
    reportNumber = "SNUTP23-002, KIAS-P23070, LCTP-23-20, SNUTP23-002; KIAS-P23070; LCTP-23-20;",
    doi = "10.1007/JHEP12(2024)029",
    journal = "JHEP",
    volume = "12",
    pages = "029",
    year = "2024"
}

@article{Chang:2024zqi,
    author = "Chang, Chi-Ming and Lin, Ying-Hsuan",
    title = "{Holographic covering and the fortuity of black holes}",
    eprint = "2402.10129",
    archivePrefix = "arXiv",
    primaryClass = "hep-th",
    month = "2",
    year = "2024"
}

@article{Chang:2024lxt,
    author = "Chang, Chi-Ming and Chen, Yiming and Sia, Bik Soon and Yang, Zhenbin",
    title = "{Fortuity in SYK models}",
    eprint = "2412.06902",
    archivePrefix = "arXiv",
    primaryClass = "hep-th",
    doi = "10.1007/JHEP08(2025)003",
    journal = "JHEP",
    volume = "08",
    pages = "003",
    year = "2025"
}

@article{deMelloKoch:2024pcs,
    author = "de Mello Koch, Robert and Kim, Minkyoo and Kim, Seok and Lee, Jehyun and Lee, Siyul",
    title = "{Brane-fused black hole operators}",
    eprint = "2412.08695",
    archivePrefix = "arXiv",
    primaryClass = "hep-th",
    reportNumber = "SNUTP24-004",
    doi = "10.1007/JHEP07(2025)216",
    journal = "JHEP",
    volume = "07",
    pages = "216",
    year = "2025"
}

@article{Chang:2025rqy,
    author = "Chang, Chi-Ming and Lin, Ying-Hsuan and Zhang, Haoyu",
    title = "{Fortuity in the D1-D5 system}",
    eprint = "2501.05448",
    archivePrefix = "arXiv",
    primaryClass = "hep-th",
    month = "1",
    year = "2025"
}

@article{Chen:2025sum,
    author = "Chen, Yiming",
    title = "{Fortuity with a single matrix}",
    eprint = "2511.00790",
    archivePrefix = "arXiv",
    primaryClass = "hep-th",
    month = "11",
    year = "2025"
}

@article{Kim:2025vup,
    author = "Kim, Seok and Lee, Jehyun and Lee, Siyul and Oh, Hyunwoo",
    title = "{BPS phases and fortuity in higher spin holography}",
    eprint = "2511.03105",
    archivePrefix = "arXiv",
    primaryClass = "hep-th",
    month = "11",
    year = "2025"
}

@article{Cabo-Bizet:2018ehj,
    author = "Cabo-Bizet, Alejandro and Cassani, Davide and Martelli, Dario and Murthy, Sameer",
    title = "{Microscopic origin of the Bekenstein-Hawking entropy of supersymmetric AdS$_{5}$ black holes}",
    eprint = "1810.11442",
    archivePrefix = "arXiv",
    primaryClass = "hep-th",
    doi = "10.1007/JHEP10(2019)062",
    journal = "JHEP",
    volume = "10",
    pages = "062",
    year = "2019"
}

@article{Choi:2018hmj,
    author = "Choi, Sunjin and Kim, Joonho and Kim, Seok and Nahmgoong, June",
    title = "{Large AdS black holes from QFT}",
    eprint = "1810.12067",
    archivePrefix = "arXiv",
    primaryClass = "hep-th",
    reportNumber = "SNUTP18-005, KIAS-P18097",
    month = "10",
    year = "2018"
}

@article{Benini:2018ywd,
    author = "Benini, Francesco and Milan, Elisa",
    title = "{Black Holes in 4D $\mathcal{N}$=4 Super-Yang-Mills Field Theory}",
    eprint = "1812.09613",
    archivePrefix = "arXiv",
    primaryClass = "hep-th",
    reportNumber = "SISSA 56/2018/FISI",
    doi = "10.1103/PhysRevX.10.021037",
    journal = "Phys. Rev. X",
    volume = "10",
    number = "2",
    pages = "021037",
    year = "2020"
}

@article{Romelsberger:2005eg,
    author = "Romelsberger, Christian",
    title = "{Counting chiral primaries in N = 1, d=4 superconformal field theories}",
    eprint = "hep-th/0510060",
    archivePrefix = "arXiv",
    doi = "10.1016/j.nuclphysb.2006.03.037",
    journal = "Nucl. Phys. B",
    volume = "747",
    pages = "329--353",
    year = "2006"
}

@article{Kinney:2005ej,
    author = "Kinney, Justin and Maldacena, Juan Martin and Minwalla, Shiraz and Raju, Suvrat",
    title = "{An Index for 4 dimensional super conformal theories}",
    eprint = "hep-th/0510251",
    archivePrefix = "arXiv",
    doi = "10.1007/s00220-007-0258-7",
    journal = "Commun. Math. Phys.",
    volume = "275",
    pages = "209--254",
    year = "2007"
}

@article{LSscft,
    author = "Choi, Jaehyeok and Kim, Seunggyu",
    title = "{Fortuity and relevant deformations}",
    journal = "to appear"
}

@article{conifold,
    author = "Jaehyeok Choi and Sunjin Choi and Seok Kim",
    title = "{New cohomologies on the conifold}",
    journal = "in preparation"
}

@article{seiberg,
    author = "Gadde, Abhijit and Lee, Eunwoo and Rajat, Raj and Tomar Shivansh",
    title = "{$Q$-cohomologies under Dualities in Gauge Theories}",
    journal = "to appear"
}

@article{Beisert:2004ry,
    author = "Beisert, Niklas",
    title = "{The Dilatation operator of N=4 super Yang-Mills theory and integrability}",
    eprint = "hep-th/0407277",
    archivePrefix = "arXiv",
    reportNumber = "AEI-2004-057",
    doi = "10.1016/j.physrep.2004.09.007",
    journal = "Phys. Rept.",
    volume = "405",
    pages = "1--202",
    year = "2004"
}

@article{Budzik:2023xbr,
    author = "Budzik, Kasia and Gaiotto, Davide and Kulp, Justin and Williams, Brian R. and Wu, Jingxiang and Yu, Matthew",
    title = "{Semi-chiral operators in 4d $ \mathcal{N} $ = 1 gauge theories}",
    eprint = "2306.01039",
    archivePrefix = "arXiv",
    primaryClass = "hep-th",
    doi = "10.1007/JHEP05(2024)245",
    journal = "JHEP",
    volume = "05",
    pages = "245",
    year = "2024"
}

@article{Budzik:2022mpd,
    author = "Budzik, Kasia and Gaiotto, Davide and Kulp, Justin and Wu, Jingxiang and Yu, Matthew",
    title = "{Feynman diagrams in four-dimensional holomorphic theories and the Operatope}",
    eprint = "2207.14321",
    archivePrefix = "arXiv",
    primaryClass = "hep-th",
    doi = "10.1007/JHEP07(2023)127",
    journal = "JHEP",
    volume = "07",
    pages = "127",
    year = "2023"
}

@article{Bomans:2023mkd,
    author = "Bomans, Pieter and Wu, Jingxiang",
    title = "{Unravelling the Holomorphic Twist: Central Charges}",
    eprint = "2311.04304",
    archivePrefix = "arXiv",
    primaryClass = "hep-th",
    doi = "10.1007/s00220-024-05167-4",
    journal = "Commun. Math. Phys.",
    volume = "405",
    number = "12",
    pages = "290",
    year = "2024"
}

@article{Gaiotto:2024gii,
    author = "Gaiotto, Davide and Kulp, Justin and Wu, Jingxiang",
    title = "{Higher operations in perturbation theory}",
    eprint = "2403.13049",
    archivePrefix = "arXiv",
    primaryClass = "hep-th",
    doi = "10.1007/JHEP05(2025)230",
    journal = "JHEP",
    volume = "05",
    pages = "230",
    year = "2025"
}

@article{Cachazo:2002ry,
    author = "Cachazo, Freddy and Douglas, Michael R. and Seiberg, Nathan and Witten, Edward",
    title = "{Chiral rings and anomalies in supersymmetric gauge theory}",
    eprint = "hep-th/0211170",
    archivePrefix = "arXiv",
    reportNumber = "RUNHETC-2002-45",
    doi = "10.1088/1126-6708/2002/12/071",
    journal = "JHEP",
    volume = "12",
    pages = "071",
    year = "2002"
}

@article{Chang:2025mqp,
    author = "Chang, Chi-Ming and Lin, Ying-Hsuan",
    title = "{Violation of S-duality in classical $Q$-cohomology}",
    eprint = "2510.24008",
    archivePrefix = "arXiv",
    primaryClass = "hep-th",
    month = "10",
    year = "2025"
}

@article{Gadde:2025yoa,
    author = "Gadde, Abhijit and Lee, Eunwoo and Raj, Rajat and Tomar, Shivansh",
    title = "{Probing Non-Graviton Spectra in $\mathcal{N}=4$ SYM via BMN truncation and S-Duality}",
    eprint = "2506.13887",
    archivePrefix = "arXiv",
    primaryClass = "hep-th",
    month = "6",
    year = "2025"
}

@article{Argurio:2003ym,
    author = "Argurio, Riccardo and Ferretti, Gabriele and Heise, Rainer",
    title = "{An Introduction to supersymmetric gauge theories and matrix models}",
    eprint = "hep-th/0311066",
    archivePrefix = "arXiv",
    doi = "10.1142/S0217751X04018038",
    journal = "Int. J. Mod. Phys. A",
    volume = "19",
    pages = "2015--2078",
    year = "2004"
}

@article{Konishi:1983hf,
    author = "Konishi, K.",
    title = "{Anomalous Supersymmetry Transformation of Some Composite Operators in SQCD}",
    reportNumber = "CERN-TH-3732",
    doi = "10.1016/0370-2693(84)90311-3",
    journal = "Phys. Lett. B",
    volume = "135",
    pages = "439--444",
    year = "1984"
}

@article{Konishi:1985tu,
    author = "Konishi, Ken-ichi and Shizuya, Ken-ichi",
    title = "{Functional Integral Approach to Chiral Anomalies in Supersymmetric Gauge Theories}",
    reportNumber = "IFUP-TH-5-85",
    doi = "10.1007/BF02724227",
    journal = "Nuovo Cim. A",
    volume = "90",
    pages = "111",
    year = "1985"
}

\end{document}